\documentclass[runningheads]{llncs} 
\usepackage[T1]{fontenc} 
\usepackage{graphicx} 
\usepackage{hyperref}
\usepackage{color} 
 
\begin{document} 
\title{Ablation and the Meno: \\ Tools for Empirical Metamathematics} 
\titlerunning{Ablation and the Meno} 
\author{Zhengqin Fan\inst{1}\orcidID{0009-0009-1048-505X} \and Simon DeDeo\inst{1,2}\orcidID{0000-0002-5346-9393}} 
\authorrunning{Z. Fan \& S. DeDeo} 
\institute{Carnegie Mellon University, Pittsburgh PA 15213 USA \and Santa Fe Institute, Santa Fe NM 87501 USA \\ \email{zhengqinfan@cmu.edu}; \email{sdedeo@andrew.cmu.edu}; \url{https://proofsandreasons.io} \\ } 
\maketitle 

\begin{abstract} 
We present the results from \emph{Meno}, a simple autoformalizer that proves theorems in Lean by systematically exploring the space of both formal and informal proofs, and \emph{tactic ablation}, a new method for exploring mathematical creativity under constraint. We show these tools in action on simple theorems found in Terrence Tao's \emph{Analysis I}, selectively ablating solution paths associated with non-constructive proofs, and analyze the properties of the resulting population using Goedel Prover embeddings. Among other things, our analysis of this novel population reveals that they lie on low (one or two) dimensional submanifolds of the much higher-dimensional representation space, and far away from their corresponding human constructions.
 \end{abstract} 

\noindent The advent of autoformalization has led to major achievements. Mathematicians are now able to ``vibe code'' important theorems \cite{armstrong2026formalization}, while A.I. companies are devoting increasing resources to proving difficult conjectures in advanced mathematics. One exciting possibility is that we might gain a new view onto the nature of mathematics itself: a field one might call empirical meta-mathematics, complementary to the more theoretical studies of highly-constrained systems found in proof complexity theory~\cite{krajivcek2019proof}, and focused on expanding our understanding of the kind of mathematics practiced on a day-to-day basis by research mathematicians in the context of a wider ``Platonic world''~\cite{barkeshli2026artificial}.

The work presented in this paper is motivated by three inter-related goals in this new domain. First, we wish to explore, in a systematic fashion, the space of possible proofs. It has long been understood that any particular theorem we care about has many proofs, which answer to a variety of needs---a theorem usually has many more or less explanatory, succinct, elegant, or counter-intuitive proofs. This suggests that it is possible to repeatedly prove the same theorem in different ways, to learn about the highlights, nadirs, and microstructure of this space.

Second, we wish to understand the relationship between formal and informal proofs of the same theorem---what Ref.~\cite{dedeo2026correspondence} refers to as the correspondence problem. For much of mathematical history, it has been assumed that a formal proof is simply an elaboration, or ``filling in the gaps'', of an informal proof, but the advent of Generative A.I. means that this is now something that can be studied on a large scale by empirical means.

Third, we wish to understand the role of constraint in theorem-proving. Mathematicians have long been fascinated not just by the theorems they prove, but how they might be proved under constraint. This fascination ranges in seriousness from parlour games where one tries to prove a simple theorem by ruling a standard approach out of bounds, to entire philosophical schools devoted to, famously, the idea that techniques such as the law of excluded middle ought to be excluded completely~\cite{sep-intuitionism}. It includes famous achievements such as the ``elementary'' proof of the Prime Number Theorem~\cite{pnt,spencer2009elementary}, and the field of reverse mathematics~\cite{sep-reverse-mathematics}.

To undertake these investigations in an empirical fashion requires new tools, and so we present, first, our human-like formalizer, \emph{Meno}, an agentic A.I. system inspired by a recent paper by Alex Kontorovich~\cite{kontorovich2025shape} and structured in such a way as to bring the formal and informal sides of proof making into sustained, iterative contact. \emph{Meno} is not a competitor to large-scale efforts in autoformalization, but is---as in the eponymous Platonic dialogue~\cite{meno}---an experimental tool for controlled studies in empirical metamathematics. We then present ``ablation'', a systematic method for imposing constraints at the formal level. With these tools in hand, we then present first results from applying these tools to simple theorems from Terrence Tao's \emph{Analysis I}~\cite{tao}.

\section{Meno: a human-like formalizer}
\label{server-side-state-machine-orchestration}

Motivated by the quasi-formalization paradigm outlined in Kontorovich~\cite{kontorovich2025shape}, we present \emph{Meno}, which works (currently) via a Model Context Protocol (MCP) server that orchestrates LLM agents for Lean 4 proof formalization. Meno takes a theorem statement in Lean, along with (optionally) a human-readable informal proof, and attempts to provide a Lean proof of the theorem. It draws in part on the open-source {\tt lean4-skills} agent~\cite{lean4-skills} and {\tt lean4-lsp-mcp}~\cite{lean-lsp-mcp}, and can be thought of as one way to compose these, and related, skills into an autonomous loop. It works cleanly with our tactic ablation method, described in Section~\ref{ablation} below, which uses Lean metaprogramming tools to selectively remove capabailities from the underlying proof system.

Meno follows a human-like process for producing formalized mathematical proofs, passing through a series of distinct phases; each phase corresponds to a collection of intellectual tasks---for example, one phase, {\bf Decompose}, corresponds to the task of taking an informal proof and breaking it down into a small number of distinct units. The key phases in our system are {\bf Set-up} (where it reads the target and looks for available tools and lemmas), {\bf Evaluate} (where it assesses the difficulty of the target theorem in the context found by Set-up), {\bf Strategize} (where it produces an informal proof, if one is not provided), {\bf Decompose} (where it breaks down the informal proof into sub-steps), {\bf Scaffold} (where it attempts to produce {\tt sorry}-ful Lean code from these sub-steps), {\bf Implement} (where it attempts to fill in those sorries), and {\bf Iterate} (where it returns to various earlier stages if a proof is not yet forthcoming). These phases can lead back into each other; the final result, just as in ordinary human mathematics, is either a solution or (in cases where a resource constraint has run out) a ``blocked'', or quit state. Each of these stages can draw both on prompted LLM outputs, and outputs from the tools such as Lean LSP MCP.

Our current implementation of Meno relies heavily on Claude code's infrastructure, where a ``worker'' LLM is prompted with the overall scheme of the different phases, and given access to various ``skills'' depending on the phase; \emph{e.g.}, we define a semi-formal skill called {\tt /Proof-Strategy} that can be called by the system as part of the {\bf Strategy} phase. 

Our use of Claude's infrastructure is sub-optimal for two reasons. First, \emph{alignment}. While we can audit Meno's actual execution trace, we do not have guarantees that the infrastructure is, in every instance, executing the phases and making choices in the way we expect. Second, \emph{cost}. In our test-runs, Meno used Claude models, usually Haiku 4.5 or Sonnet 4.6, as the orchestrator of this process, and it is unclear whether smaller models, that operate below Haiku-level, or without the infrastructure provided by Claude, will be as successful. The effective cost per proof, even for the short proofs presented here, ranges from 50\textcent~to \$3. 

A key future goal for this system is to more cleanly separate out the deterministic high-level logic from the more non-deterministic behavior of (for example) LLM-enabled proof decomposition or {\tt sorry}-filling, so that we can selectively call lower-resource LLMs when needed. This would also give us greater control over the way in which the space is explored, and more information about failed directions. The current version of our system, expensive as it is, provides a test case for this kind of autonomous theorem proving.

\section{Constraining Proof Strategies through Tactic Ablation}
\label{ablation}

Lean is a dynamic and ongoing project, and the tactics and theorems available to its users are constantly expanding. Its community is focused on augmentation---giving users more powers---rather than philosophical or foundational hygene \cite{tpil}. This is particularly true for Lean 4, which supports a broad user-base, including many who are relatively uninterested in the performance and maintainability tradeoffs required to cleanly expose foundational distinctions to the interface.

A complete solution, that can consistently forbid a proof strategy across all occurrences is thus likely impossible. Even apparently unconnected axioms can introduce powers, that we might have wanted to remove, by the back door: the fact that the Axiom of Choice implies the Law of Excluded Middle, for example, only came to broad notice many decades after Brouwer's original presentation of intuitionism. Our goals, however, are more modest: we are interested in what happens when Meno finds some paths newly \emph{difficult}, not newly impossible. 

Our solution relies on a principled partitioning of the Lean 4 tactic vocabulary into groups that can be independently withheld---``ablated''---from an agent. Practically speaking, tactics are withheld within Lean itself; information, about what is and is not allowed, reaches Meno via its interactions with the Lean LSP: either direct interactions, or via skill calls, such as \verb|lean_multi_attempt|, that rely on LSP information. Our method also allows for a lemma ``whitelist'' to be applied; this is useful in cases where we are interested in how Meno might approach, and solve, a problem where a solution already exists in mathlib. The whitelist allows us to restrict Meno to, for example, only the mathlib lemmas that have been previously used in earlier theorems in the project.

We organize tactics along three dimensions (\emph{abstraction level}, \emph{functional category}, and \emph{axiom dependency}) so that each ablation experiment can target a semantically coherent slice of the tactic space. Our focus in this test case is selective ablation of non-constructive tactics.

\noindent \textbf{$\cdot$ Abstraction levels (vertical axis).} Each category is assigned an integer level from 0 to 4. These levels form a dependency hierarchy: higher-level tactics may internally invoke lower-level ones, but not vice versa. At Level 0 (Atomic), we have tactics that do not internally invoke other tactics, including term-level primitives (\texttt{exact}, \texttt{apply}, \texttt{intro}, \texttt{cases}, \texttt{constructor}) and control flow combinators (\texttt{first}, \texttt{repeat}, \texttt{all\_goals}). At Level 1 (Proof Orgaization), we have tactics that restructure the proof state without invoking automated reasoning (e.g.~\texttt{have}, \texttt{suffices}, \texttt{rw}). At Level 2 (Normalization), we have rewrite-rule engines and canonicalizers (e.g.~\texttt{simp}, \texttt{ring}, \texttt{norm\_num}). At Level 3 (Decision procedures and search), we have tactics backed by dedicated solvers or library-wide search (e.g.~\texttt{omega}, \texttt{linarith}, \texttt{grind}). Finally, at Level 4 we have domain-specific tactics that encode knowledge about particular mathematical domains (e.g.~\texttt{continuity}, \texttt{measurability}, \texttt{finiteness}). 

\noindent \textbf{$\cdot$ Functional categories (horizontal axis).} Each tactic is assigned to one of approximately 35 categories derived from two sources. For core Lean tactics (those defined in \texttt{Init.*}, \texttt{Std.*} and \texttt{Lean.*} modules), we adopt the section headings of the official Lean 4 Tactic Reference: for example, \emph{Assumptions}, \emph{Quantifiers}, \emph{Simplification}, and \emph{Control Flow}. For mathlib-defined tactics, we derive categories from the mathlib source module directory structure: for instance, tactics registered under \texttt{Mathlib.Tactic.Ring.*} map to \emph{Ring Normalization}.

\noindent \textbf{$\cdot$ Axiom tiers (depth axis).} Orthogonally, a companion audit classifies each tactic by the logical axioms its proofs may introduce. This classification proceeds in two stages. (1) it assigns a tier to each tactic using a static base, which is initiated by writing custom \verb|#print axioms| tests using an LLM, followed by variant inheritance (e.g.~\texttt{simp!} inherits from \texttt{simp}), and prefix-based family rules; (2) an empirical probing stage validates these assignments: each tactic is tested on six probe theorems (covering natural-number arithmetic, integer inequalities, propositional logic, simple rewrites, real-valued positivity, and existential witnesses) with \verb|#print axioms| appended below. The strongest axiom set of the successful attempts is recorded and compared against the assignment from the static base. The tier of this tactic is updated when the probed axiom set is stronger.

The three Axiom tiers are:
\begin{itemize}
\item \emph{Strongly constructive} (57\% of tactics): no axioms beyond the Calculus of Inductive Constructions. Includes kernel-level tactics such as \texttt{exact}, \texttt{apply}, and \texttt{cases}, the decision procedure \texttt{decide}, and structural combinators like \texttt{calc}. 
\item \emph{Weakly constructive} (29\%): may invoke \texttt{propext} (propositional extensionality) or \texttt{Quot.sound} (quotient soundness). Typical members are the normalization tactics \texttt{simp}, \texttt{ring}, \texttt{norm\_num}, and the decision procedures \texttt{omega}. 
\item \emph{Classical} (14\%): can introduce \texttt{Classical.choice}. This tier includes \texttt{by\_contra}, \texttt{contrapose}, \texttt{push\_neg}, \texttt{choose}, and \texttt{field\_simp}, as well as domain-specific tactics such as \texttt{continuity} and \texttt{measurability}. 
\end{itemize}


\section{Visualizing the space of possible proofs}

A natural first step, upon being presented with the output of a system like Meno, operating repeatedly on the same theorem, is to try to arrange the proofs it produces in some sort of space, to help us identify the variety of outputs and their relative similarities. 

This is a non-trivial task because a proof---whether in natural language, or in Lean---is a richly-structured object whose features answer to a wide variety of demands. Only some of these demands, in turn, actually track what we might mean when we compare, say, three proofs to say which one is the outlier.

There are syntactic, stylistic, and rhetorical constraints, for example, that may make proofs A and B look very different from each other in terms of word usage, while they might be, in some deeper sense, essentially equivalent strategies compared to a third proof, C. Simple tools from NLP, such as word counts or topic modeling, will struggle to cleanly separate out the essential from the inessential.

In Lean, of course, it is always possible to expand code until one reaches the directed acyclic graph that replicates the core, foundational type-theoretic concepts, as was done, for example, in Ref.~\cite{viteri2022epistemic}. It is still unclear, however, how to take  DAGs produced in this fashion and determine how similar, or how different, they ``truly'' are at a semantic level. One can try to use summary statistics (how many nodes, how many leaves, in- and out-degrees, and so forth) but very different proofs---as Ref.~\cite{viteri2022epistemic} showed---often have very similar graph properties. 

Even if we include semantic information about the nodes (\emph{e.g.}, the names of the axioms used), a great deal of hidden variation---and a great deal of irrelevant surface variation---is possible if we restrict to summary statistics. Conversely, the extreme size of these DAGs (a proof of Russell's Paradox, for example, may take a few dozen words of elementary Lean, but expand to many hundreds of nodes and deges) makes simple, but more principled, ideas, such as graph edit distances, essentially impossible to implement.

Our solution uses Goedel-Prover-V2-32B~\cite{lin2025goedel}, a large language model fine-tuned to produce Lean code. Goedel Prover's internal activations can be analyzed, with reasonable amounts of computer power, to study the representations it produces in the course of ``reading'' a proof. Empirically, we know that the activation states of a neural network that does next-token prediction correspond, roughly, to the system's internal representation of the context in which that token is produced. This empirical insight is the basis of ``text embeddings'' that are a powerful tool for natural language processing~\cite{opitz2025interpretable}. Using embeddings has the further advantage---when it works---of mapping a complex object (a natural language text or, in our case, highly-structured Lean code) into a simpler one (a vector space, e.g., $R^n$, where $n$ might be large, but all the standard tools, such as Euclidean distance, become available).

When doing semantic analysis in this fashion, there are two decisions to be made. The first is which layer to examine: at the very final layer, we capture only the probability distributions over the next token---the fact, for example, that the next thing to write is almost certainly ``Q.E.D'', but not the residual trace of how the system got there or knows that this is what to say next. There is little systematic knowledge about layer archaeology, but empirical experience suggests that as the ``neuroscientist'' goes deeper (\emph{i.e.}, earlier) in the layers of the neural network, she captures more and more semantic information---until, past some middle point, this semantic information is gradually and in turn replaced by surface features of the original context.

The second decision is how to \emph{combine} representations within the text; informally, if the embedding captures the ``state of mind'' of the system at a particular point (and the first decision is to ask where that can best be found), the second decision amounts to deciding what to retain of that state of mind over the course of the proof-reading.

As a first pass at answering these two questions, we (1) sample the network at layer 25, roughly mid-way between the input (layer 0) and final prediction layer 63) of Goedel Prover's Qwen-style architecture,\footnote{Layer 25 leads to pairwise proof-proof distance distributions with the highest kurtosis, maximizing structure---some pairs are close, but separated from others. Not much hangs on the exact choice of layer, however, and we recover similar results for a range of choices between zero and 63.} and (2) use the embedding at the final token of the proof; the proof includes the type signature, but is stripped of any comments. The embedding vector is 5140 dimensions.

The next step of this process is visualization; we use a classic algorithm, Multi-Dimensional Scaling (MDS~\cite{mead1992review}) to project down this very high dimensional space, on the basis of the between-proof distances, into two dimensions in such a way that the underlying distance information is preserved as well as possible. This makes it possible to display, on the page, a summary map of what the space of proofs looks like. MDS is not the only possible tool for dimensionality reduction---other methods, such as UMAP and tSNE, have similar goals, and more complex algorithms, but our tests on our sample theorems suggest that MDS performs better at retaining the original distance information. 

To aid visualization, we run a Bayesian clustering model (the Gaussian Mixture Model), which attempts to represent the points in the reduced MDS space as a collection of Gaussian blobs, with arbitrary orientation, sizes, and aspect ratios. The final product of this analysis is a first representation of the space of proofs of the theorem in question, made visually tractable.

\section{Results}

As a test case and proof of concept for the Meno and Ablation methods, we use three theorems from Tao's \emph{Analysis I}~\cite{tao}: {\tt Russells\_paradox}, {\tt EqualCard.} {\tt power\_set\_false} ($|X| < \left|2^X\right|$), and {\tt Sequence.Cauchy\_iff\_convergent} (completeness of the reals). A whitelist prevents Meno from accessing the mathlib versions of these theorems, and restricting it those used by Tao's Lean code up to that point in the text. Meno repeatedly proved all three theorems, under two conditions: free (whitelist only), and constructive (whitelist+ablation of {\tt Classical.choice}). 

\begin{figure}
    \centering
    \includegraphics[width=0.9\linewidth]{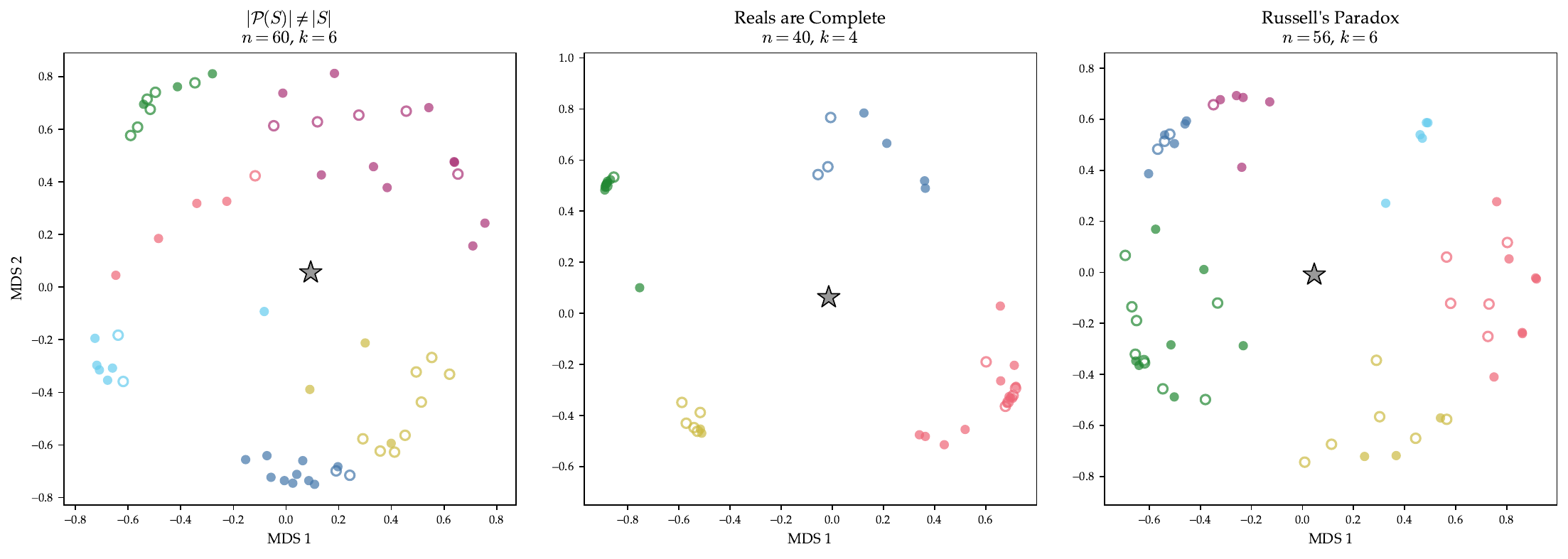} \\
    \includegraphics[width=0.9\linewidth]{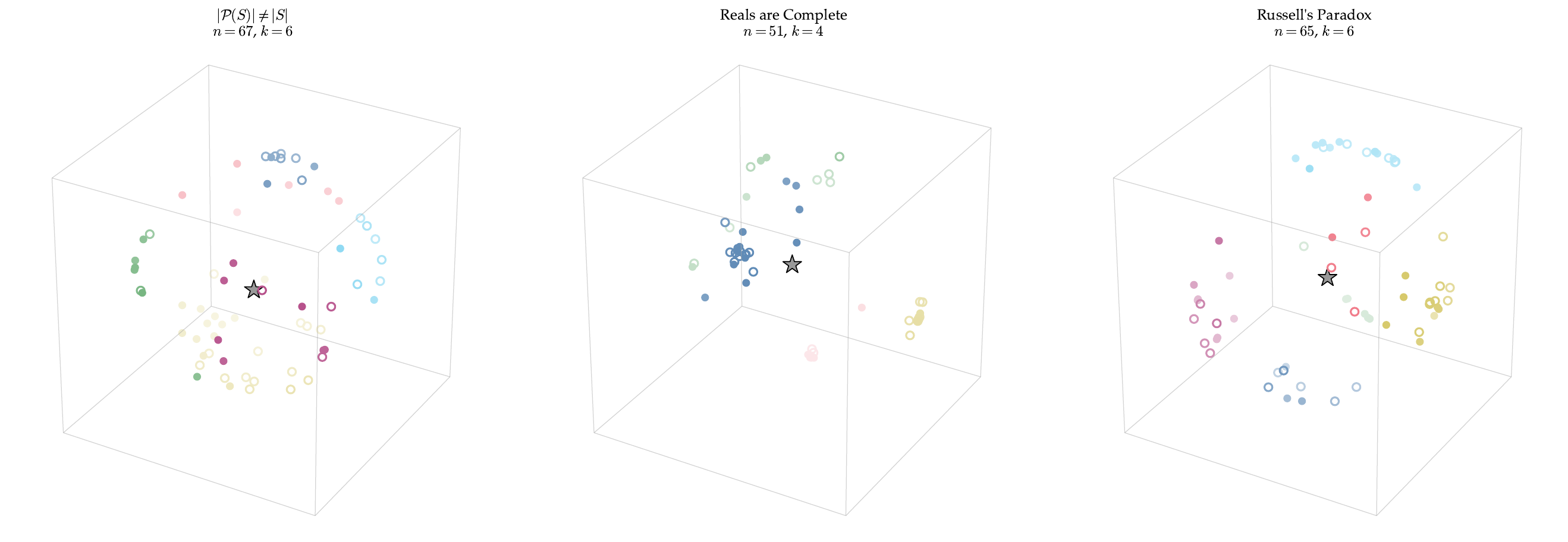}
    \caption{The space of proofs as found by Meno, both with and without ablation (open vs.\ solid circles), for the three proofs of Table~\ref{counts}. Tao's Lean proof is marked as a star.}
    \label{cauchy}
\end{figure}

\begin{table}[]
    \centering
    \begin{tabular}{r|c|c|c|c|c}
Theorem~ & ~Plain ($n$)~ & ~Ablated ($n$)~ & ~1d MDS $r$~ & ~2d MDS $r$~ & ~3d MDS $r$ \\
\hline
Russell's Paradox~ & 36 & 29 & 0.289 & 0.912 & 0.941 \\
$|X| < |2^X|$~ & 36 & 31 & 0.353 & 0.876 & 0.913 \\
Reals are complete~ & 25 & 26 & 0.536 & 0.965 & 0.989 \\
     \end{tabular}
    \caption{The proofs we analyze, by type (plain, or ablated of {\tt Classical.choice}), and the correlation between distances in the 5140 dimensional embedding space and the dimensionality-reduced MDS projection. Remarkably, nearly all of the distance information can be captured by low dimensional embeddings.}
    \label{counts}
\end{table}
Table~\ref{counts} shows the correlations between the embedding-space cosine distances, and the one-, two-, and three-dimensional distances found by MDS. The first, remarkable result of this analysis is that, for all three theorems, these proofs lie naturally in a much lower dimensionality space; nearly all of the variation can be captured with three diemsions. Fig.~\ref{cauchy} shows how these proofs actually arrange themselves in MDS space, with Tao's original Lean proof noted as a star. Meno's proof strategies not only live in low-dimensional space but are often constrained, within that space, to complex and twisted, but roughly one- or two-dimensional submanifolds. 

In all three cases, the human proof (Tao's Lean) lies quite far from all these submanifolds; effectively equidistant from all the other points, it is placed by MDS in the center. Intriguingly, ablation, for these simple theorems, was not an overriding driver of proof location in embedding space; it is easy to find non-ablated (filled circle) proofs nearby ablated ones. This may be in part because the natural proofs Meno finds are often, themselves, constructive, while in other cases, such as Russell's Paradox, the constructive proof is, thanks to Lean's type theoretic orientation, very similar to the non-constructive one.

\section{Conclusion}

The main goal of this work was to create, and test, Meno, a system for metamathematical exploration that uses natural language proofs to guide the construction of formal proofs in Lean, and that can be selectively ``ablated'' to constrain the strategies and prior lemmas it can use

Meno's outputs provide a first view onto the landscape of mathematical proofs, and the distinctive strategies they use and that separate them from the human case. Perhaps most surprising is theie dimensionality; despite the variety of surface-level syntactical choices, proofs form low-dimensional structures and arrange themselves in simple clusters, lines, and two-dimensional shapes.

\begin{credits} \subsubsection{\ackname} This work was supported by Grant 63750, “Explaining Universal Truths”, from the John Templeton Foundation, and Allocation MTH260032 at Bridges-2 at Pittsburgh Supercomputing Center, via the ACCESS program (NSF grants \#2138259, \#2138286, \#2138307, \#2137603, and \#2138296). We thank Dietrich Computing \& Operations for support, and Max Noichl and David Barack for feedback.

\subsubsection{\discintname} The authors have no competing interests to declare. \end{credits} 
%
%

\bibliographystyle{splncs04} 
\bibliography{references}

\end{document}